# Mastering hysteresis in magnetocaloric materials

O. Gutfleisch*, T. Gottschall, M. Fries, D. Benke, I. Radulov, K. P. Skokov

*Materialwissenschaft, Technische Universität Darmstadt, Alarich-Weiss-Str. 16, 64287 Darmstadt, Germany,*

H. Wende, M. Gruner, M. Acet, P. Entel and M. Farle

*Fakultät für Physik, Universität Duisburg-Essen, Geibelstr. 41, 47057 Duisburg, Germany*

## Summary

Hysteresis is more than just an interesting oddity, which occurs in materials with a first-order transition. It is a real obstacle on the path from existing lab-scale prototypes of magnetic refrigerators towards commercialization of this potentially disruptive cooling technology. Indeed, the reversibility of the magnetocaloric effect, being essential for magnetic heat pumps, strongly depends on the width of the thermal hysteresis and therefore it is necessary to understand the mechanisms causing hysteresis and to find solutions how to minimize losses associated with thermal hysteresis in order to maximize the efficiency of magnetic cooling devices. In this work, we discuss fundamental aspects, which can contribute to thermal hysteresis and we are developing strategies for at least partially overcoming the hysteresis problem in some selected classes of magnetocaloric materials with large application potential. Doing so, we refer to the most relevant classes of magnetic refrigerants La-Fe-Si-, Heusler- and $Fe_2P$-type compounds.

## Introduction

Magnetic refrigeration technology is a fast-developing technology, which is assumed to be capable to compete with and hopefully surpass the traditional vapour-compression refrigeration in terms of efficiency, device volume and ecological impact in the near future [2]. From this point of view, magnetic compounds exhibiting large magnetocaloric effects (MCE) in the temperature range of 270 – 320 K (under magnetic field changes of $\Delta(\mu_0 H)$ = 1 - 2 T) attract much attention due to their potential application in magnetic refrigeration at room temperature [3-7]. The search for new magnetocaloric materials as well as the optimization of known materials is a growing field of interest, and the new magnetic materials showing a large magnetocaloric effect (MCE) with narrow hysteresis, particularly around room temperature, are the subject of intensive research activity. A variety of new materials have been synthesized since the discovery of the first giant magnetocaloric material $Gd_5(Si,Ge)_4$ in 1997 [8, 9], and some of them have proven to be very promising for applications. The main issue related to giant MCE materials is the hysteresis of the first-order transition which is one of the main factors delaying the development of this novel and disruptive cooling technology. This paper aims (a) to summarize the fundamental phenomena which can contribute to thermal and magnetic hysteresis, and (b) to develop strategies for at least partially overcoming or by passing the hysteresis problem in some selected classes of magnetocaloric materials with large application potential.

The magnetocaloric effect itself can be defined as the reversible change of the thermodynamic variables of the sample - temperature $T$ and entropy $S$ – caused by a variation of the applied external magnetic field. If the magnetization or demagnetization of the sample is performed under adiabatic conditions, it results in a heating or cooling of the material. In this case, the total entropy $S$ remains constant and the magnetocaloric effect manifests itself in an adiabatic temperature change $\Delta T_{ad}$. Alternatively, by keeping the temperature $T$ constant during magnetization or demagnetization (isothermal conditions), the magnetocaloric effect leads to a transfer of thermal energy between the sample and the environment. The amount of heat $Q$ transferred in this process can be quantified as $Q = T\Delta S_m$ where $\Delta S_m$ is the isothermal magnetic entropy change.

An overwhelming majority of researchers investigate the MCE by measuring the isothermal $M(H)_T$ or isofield $M(T)_H$ magnetization with subsequent recalculation (the indirect method) of $\Delta S_m(T)$ by using the Maxwell relation. As a result, most of the published MCE data are temperature dependences of magnetic entropy changes $\Delta S_m(T)$ obtained in this way. The $\Delta S_m$ is an important parameter, but at the same time, a high value of $\Delta S_m$ is not sufficient for a material to be a good magnetic refrigerant [10, 11]. Furthermore, the integration of the $\Delta S_m(T)$ curve (sometimes referred to as the relative cooling power - RCP=$\Delta S_m \, \partial T_{FWHM}$ [7]) is usually misleading [12]. To assess the magnetocaloric properties adequately, the adiabatic temperature change $\Delta T_{ad}$ needs to be taken into account as well. However, experimental data on $\Delta T_{ad}$ are rather scarce in literature.

†Present address: Material Science, Technische Universität Darmstadt, Alarich-Weiss-Str. 16, 64287 Darmstadt, Germany
*Author for correspondence (gutfleisch@fm.tu-darmstadt.de).

Thus, by using both $\Delta T_{ad}$ and $\Delta S_m$, it is possible to comprehensively characterize magnetocaloric materials in terms of their potential for magnetic refrigerators. Both $\Delta T_{ad}$ and $\Delta S_m$ can be described within the framework of the *S(T)* diagram, which can be plotted by using temperature dependences of heat capacity $c_p$ measured in different magnetic fields. The conventional *S(T)* diagram describes the thermal equilibrium situation and it does not take into account metastability and hysteresis of materials with first-order transition. Moreover, this thermal equilibrium would be achieved only after very long duration. Since in a magnetic refrigerator the magnetocaloric material should be magnetized and demagnetized very fast, standard calorimetry followed by subsequent analysis of the equilibrium *S(T)* diagram cannot mirror these operational conditions. Furthermore, we found that the *S(T)* diagram obtained under continuous heating and cooling fails to describe the reversible magnetocaloric properties of minor loops under cycling [13]. At the same time, cyclic measurements of the adiabatic temperature change together with calorimetric data allow to determine the reversible magnetic field induced entropy change $\Delta S_m$ in fast operation comparable to real device conditions which can, in principle, be applied to every magnetocaloric material with a first-order transition.

Along with $\Delta S_m$ and $\Delta T_{ad}$, the third primary parameter which characterizes the suitability of the magnetocaloric material is the thermal conductivity $\lambda = \alpha \rho c_p$, where $\alpha$ is the thermal diffusivity and $\rho$ the density of the material. It determines the ability of the material to transfer thermal energy to the heat exchange fluid. The thermal conductivity is one factor limiting the maximum operation frequency of the magnetocaloric material and therefore has to be considered together with $\Delta S_m$ and $\Delta T_{ad}$ when discussing the applicability and efficiency of the material as a magnetic refrigerant [14-19].

In order to use the magnetocaloric material in a magnetic refrigerator, it should on the one hand be machined into a heat exchanger with a fine porous structure, designed to provide the largest possible contact surface to the heat transfer liquid. On the other hand, the effective volume of the MCE heat exchanger needs to be maximized for an efficient utilization of the (Nd-Fe-B) permanent magnet field source. The magnetic system is the most costly part of the device, and it has the highest negative environmental footprint as shown in a life cycle analysis of a magnetic refrigerator [20, 21] and therefore should be used as economically as possible. Another significant issue during the machining process of the magnetocaloric material into plates or porous structures is the possible reduction of its magnetocaloric properties [22]. As a result, a material with a high magnetocaloric effect in bulk form can show only a modest MCE after it has been shaped into a heat exchanger with sub-millimeter channels [23, 24].

## First- and second-order transitions, conventional and inverse magnetocaloric effect.

In Figure 1(a), a schematic of the magnetic behavior near the Curie temperature $T_C$ of a ferromagnetic material is illustrated. The green curve represents *M(T)* in zero field. At $T_C$, the magnetization vanishes and the ferromagnet turns into a paramagnet. This purely magnetic transition is a thermodynamic transformation of second-order. The element Gd is one of the most prominent materials undergoing such a transformation near room temperature [25]. It is worth noting that the zero-field magnetization curve cannot be measured directly due to the formation of magnetic domains, but it can be calculated for instance by Kuz'min's approach [26]. In magnetic fields, a certain magnetization is also observed in the paramagnetic phase due to the partial alignment of the spins (orange curve in Figure 1 (a)). The magnetic susceptibility of a paramagnet $\chi$, being the ratio of the magnetization *M* and the magnetic field *H*, decreases with temperature, which can be described by the Curie-Weiss law [27]:

$$\chi = \frac{M}{H} = \frac{C}{T - T_C} ,$$

with *C* being the Curie constant. Figure 1 (b) shows the entropy of the material as a function of temperature in the vicinity of the second-order transition. The total entropy is a combination of three terms, namely the contribution of the lattice $S_{lat}$, the magnetic moments $S_{mag}$ and the electronic contribution $S_{el}$ resulting in [3]

$$S(T,H) = S_{lat}(T,H) + S_{mag}(T,H) + S_{el}(T,H).$$

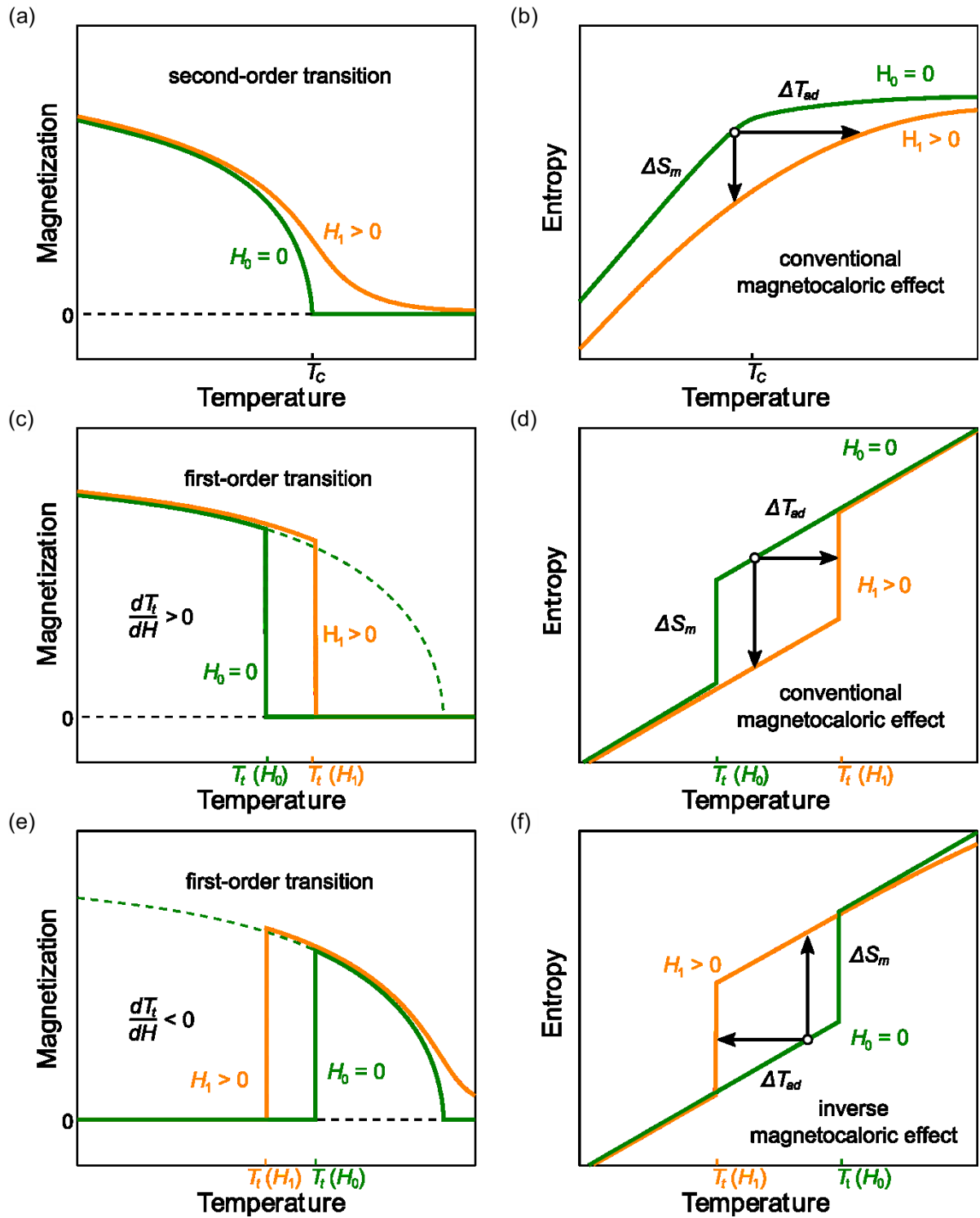

**Figure 1: Schematic of the temperature-dependence of magnetization and the total entropy with and without a magnetic field of a conventional second-order transition in (a) and (b), of a conventional first-order transformation in (c) and (d), and of an inverse first-order transition in (e) and (f).[1]**

The temperature dependence of the total entropy without magnetic field is schematically shown as a green curve in Figure 1 (b). By applying a magnetic field, the magnetic moments align to some extent. Therefore, the order of the magnetic system increases, which relates to a decrease of the magnetic entropy. For this reason, the external magnetic field lowers the entropy, which is illustrated by the orange curve in Figure 1 (b). This decrease in entropy is observed under isothermal conditions. The respective entropy change $\Delta S_m$ is plotted as a vertical arrow. Under adiabatic conditions, the total entropy stays constant. In order to compensate the decrease in the magnetic contribution, the lattice entropy increases. For this reason, the application of a magnetic field results in a heating of the material by $\Delta T_{ad}$, which is illustrated as a horizontal arrow.

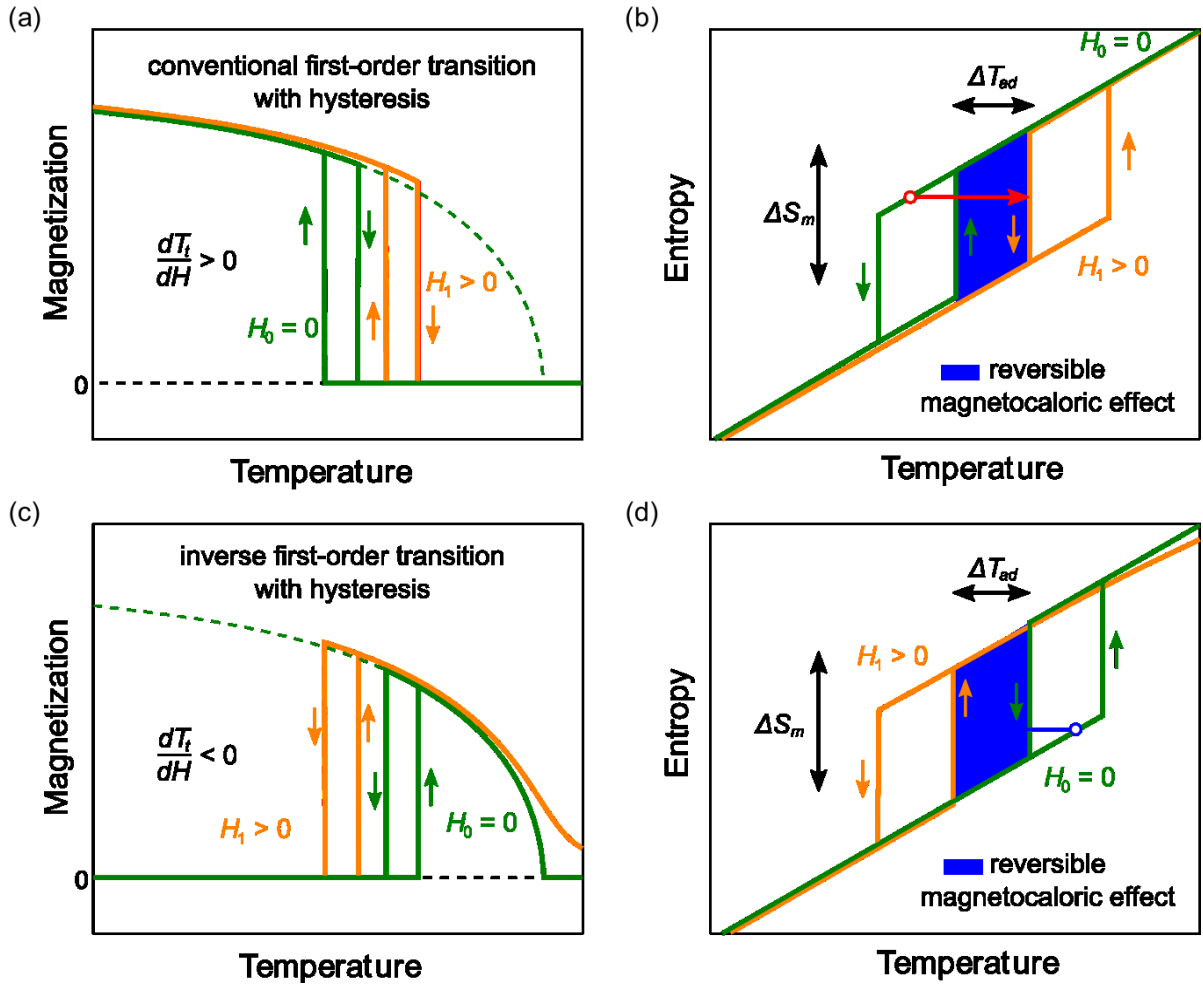

**Figure 2**: Schematic illustration of the temperature dependence of magnetization and entropy of a conventional ((a) and (b)) and an inverse first-order transition considering thermal hysteresis.[1]

In contrast to that, a conventional first-order transition is schematically shown in Figure 1 (c,d). Such a transformation is for instance observed in the material La-Fe-Si [28]. A discontinuity in magnetization is observed in both, the magnetic and total entropy curves, due to the transformation between two different phases. In the idealized picture in Figure 1(c), the ferromagnetic phase is stable at low temperatures. At $T_t$, the material transforms into the high temperature phase with low magnetization. In principle, the low temperature phase would still be ferromagnetic up to its Curie temperature, which is illustrated by the extrapolated magnetization curve (green dashed line). However, this is prevented by the magnetostructural or magnetoelastic transition.

The application of a magnetic field results in the shift of the transition temperature $T_t$. This is happening because the magnetic field stabilizes the phase with higher magnetization, being the low temperature phase [29]. The shift of the transition temperature in magnetic fields $dT_t/dH$, which is positive for a conventional first-order transition, can be understood as the driving force of the magnetocaloric effect in such a material. For instance, if one would stay in the paramagnetic phase at a temperature between $T_t$ ($H_0$) and $T_t$ ($H_1$), the application of the magnetic field $H_1$ would result in the conversion of the material into the low temperature phase. In terms of the magnetocaloric effect, it now depends on whether isothermal or adiabatic conditions are present.

The corresponding total entropy diagram is shown in Figure 1 (d). Also in the entropy curve, a discontinuity is visible in the ideal case. However, due to the shift of the transition temperature in magnetic fields, the $S(T)$ diagram has the shape of a parallelogram (green and orange curve in Figure 1 (d) [10]. When keeping the temperature constant, the entropy decreases under field application. Therefore, $\Delta S_m$ is also negative (vertical arrow in Figure 1 (d) as it was the case for the second-order transition above. When a magnetic field is applied adiabatically, the $S(T)$ diagram is crossed horizontally, which results in an increase in the temperature of the material. It is worth noting that in the schematic, the entropy curves do not overlap in the low temperature region of Figure 1(c,d). This is because the magnetic field affects the ferromagnetic phase by slightly increasing the ordering of the magnetic moments, counteracting the disorder due to thermal fluctuations. Therefore, the magnetization is slightly increased as shown by the difference between the green and the orange curve in Figure 1 (c). Consequently, the total entropy is slightly reduced in the ferromagnetic phase under magnetic field application (also the case for the paramagnetic phase but less pronounced). A similar effect iss also observed in Figure 1(a,b), which was related to the purely second-order transition. This implies that in Figure 1 both, the first- and the second-order transition overlap.

As a third example, the inverse first-order transformation is schematically illustrated in Figure 1 (e,f). Such a transition is observed for instance in Ni-Mn-X-Heusler alloys with X = Sn, Sb and In [30] or in Fe-Rh [31]. In the inverse case, also a structural phase conversion is taking place. In contrast to Figure 1 (c,d), the low temperature phase has a low and the high temperature phase has a high magnetization, respectively (see Figure 1(e)). For this reason, a magnetic field shifts the transition to lower temperatures and $dT_t/dH$ is negative. The corresponding, total entropy diagram is shown in Figure 1(f). In the ideal case, the $S(T)$ diagram has the shape of a parallelogram too, but in comparison to Figure 1(d), the green and the orange curve are exchanged. As a result, the adiabatic temperature change $\Delta T_{ad}$ is negative and the isothermal entropy change $\Delta S_m$ is positive. Again, the first and the second-order transition are overlapping, which can be seen by the increased magnetization in magnetic fields in Figure 1 (e) and by the crossing of the green and the orange $S(T)$ curve in Figure 1(f). It is worth noting that the both curves come together again for temperature well above the Curie point.

The situation changes with magnetocaloric materials of first-order which show a thermal hysteresis. This means that the back and forth transformation does not take place at the same temperature. In fact, the heating and the cooling branch of the magnetization curve are separated by the thermal hysteresis, which is illustrated for a conventional first-order transition in Figure 2 (a). The corresponding $S(T)$ diagram is shown in Figure 2(b). Such a behavior is for instance observed in certain $Fe_2P$-type materials or in La-Fe-Si [32]. Due to the existence of the thermal hysteresis, also the entropy curves under heating and cooling do not coincide. This has far-reaching consequences on the magnetocaloric effect under magnetic-field cycling. If there would be no thermal hysteresis related to the transition, then the $\Delta T_{ad}$ and $\Delta S_m$ as shown in Figure 1 (a-f) would be the same in cyclic operation. But due to the hysteresis, a reversible effect can only be obtained in the highlighted area.

The red arrow in Figure 2 (a,b) illustrates how the material being in the purely high temperature phase (after cooling from temperatures well above the transition) would heat when the magnetic field is applied the first time. About half of the material would transform in this example. However, when the magnetic field is removed again, the initial state could not be reached. In fact, the adiabatic temperature change under cycling would be smaller as indicated by the horizontal double-sided arrow. This description of the thermal hysteresis can be transferred straightforwardly to the first-order transition of inverse magnetocaloric materials, which is shown in Figure 2(c,d).

## Magnetic refrigeration and magnetocaloric materials

The development of a magnetocaloric cooling system based on the magnetocaloric principle at ambient temperature has not yet been realized as an economic and industrial product. Since the first demonstrator developed by Brown in 1976 [33] , only a few research groups around the globe have worked on demonstrators operating close to ambient temperature [34-39]. There are a number of problems to be solved, most importantly to find suitable materials with large MCE in small fields using non-critical materials and efficient heat exchangers operating at high frequencies. Recently, new but yet to be validated design concepts have been proposed; e.g. the combination of the electro- and the magnetocaloric effect [40] or the utilization of the thermal-diode mechanism [41]. Till now, a coordinated research efforts involving engineers tackling design issues of the machine on the one hand and material scientists (theory and experiment) performing fundamental research on the other hand has not been realized.

There are four main issues that need to be solved before the technology of magnetic refrigeration can move forward from prototypes to mass production.
1) The availability of low-cost, low-hysteresis magnetic refrigerants, which can provide a substantial cooling power in magnetizing fields being small enough to be provided by permanent magnets (e.g. Nd-Fe-B), rather than superconducting- or electro-magnets.
2) The price of the magnetic system, made from permanent magnets creating a magnetic field change of ideally 0.5 – 1.2 T, needs to be reduced significantly [35].
3) The switching of the heat exchange fluid should be drastically accelerated and the valves should be able to commutate the hot and cold heat exchange fluid at frequencies of 10 Hz or higher [42].
4) The time of heat transfer from the magnetocaloric material to the liquid should be reduced down to 50 – 20 ms and in this time scale the heat exchange should have 80 % – 90 % of the theoretical efficiency [43].

This last issue requires magnetocaloric materials with outstanding MCE, excellent mechanical integrity,

chemical stability and high thermal conductivity, shaped into a porous structure with fine and straight channels.

Currently, the materials utilized as magnetic refrigerants in existing prototypes of magnetic refrigerators can be divided in the following classes based on their hysteresis width [36]. In order to improve the comparability of the magnetocaloric properties of the different materials, we linearly interpolate the given literature values of the adiabatic temperature change and the isothermal entropy change to a magnetic field change of $\Delta(\mu_0 H) = 1$ T . It is worth noting that a certain error is associated with this approach since the scaling of the magnetocaloric effect is not necessarily linear [44, 45].

The first of them (non-hysteretic materials with second-order transition like in Figure 1(a,b)) includes elemental Gd and Gd-based rare-earth alloys, such as Gd-Y, Gd-Tb etc. Pure Gd metal is the benchmark magnetocaloric material, exhibiting an isothermal entropy change of $\Delta S_m \approx 3$ Jkg$^{-1}$K$^{-1}$ and an adiabatic temperature change of $\Delta T_{ad} \approx 2.5$ K under $\Delta(\mu_0 H) = 1$ T [25]. This metal is rather ductile and can be easily machined into heat exchangers [33, 46]. However, this rare-earth metal is critical in terms of its resource efficiency and environmental footprint and would be quickly too expensive when used in large amounts assuming that the market entry and mass production of magnetic refrigerators will be achieved [21, 47].
La$_{0.67}$Ca$_{0.33-x}$Sr$_x$MnO$_3$ (LCSM) manganites with a second-order transition (Figure 1(a,b)) have a low cost and good corrosion resistance, but their MCE is rather modest. Under magnetic field change of $\Delta(\mu_0 H) = 1$T, they exhibit $\Delta T_{ad} \approx 1.5$ K and $\Delta S_m \approx 5$ Jkg$^{-1}$K$^{-1}$ [48].

The second class (first-order materials with narrow thermal hysteresis Figure 1(c,d)) includes La(Fe,Si)$_{13}$-based alloys and Fe$_2$P-type compounds like Mn-Fe-P-Si. Being much more abundant and thus much cheaper than Gd metal, these materials are very attractive for commercial usage. With respect to the substitution metal used to bring the alloy's transition temperature to room temperature, it can be divided in two subclasses: La(Fe,Co,Si)$_{13}$ and La(Fe,Mn,Si)$_{13}$H$_x$. Under magnetic field change of $\Delta(\mu_0 H) = 1$ T, despite the low value of $\Delta T_{ad} \approx 1.5$ K, La(Fe,Co,Si)$_{13}$ materials demonstrate rather high $\Delta S_m \approx 5$ Jkg$^{-1}$K$^{-1}$ [49-51]. Nevertheless, these materials are brittle and it is necessary to use a special processing route in order to bring this material into the required form [23, 52-54].
La(Fe,Mn,Si)$_{13}$H$_x$-based alloys show attractive magnetocaloric properties under $\Delta \mu_0 H = 1$T: $\Delta S_m \approx 10$ Jkg$^{-1}$K$^{-1}$ and $\Delta T_{ad} \approx 3.5$ K [14, 28, 55]. However, after hydrogenation these alloys exist only in powder form due to the inevitable decrepitation of the initial bulk alloy. This powder can be embedded in a polymer-binder matrix and can for example be shaped into a heat exchanger with certain geometry. Conversely, the epoxy dilutes the MCE material and the properties of the composite can be significantly different in comparison to pure La(Fe,Mn,Si)$_{13}$H$_x$ powder as reported in literature [24, 55, 56].

The Mn-Fe-P-Si compounds are also regarded as promising for application of magnetic refrigeration at room temperature, but they are still not widely used in prototypes. Their thermal conductivity is rather low, which in turn limits the maximal operation frequency of the Fe$_2$P -type materials as magnetic refrigerants [14, 18]. In these compounds, $\Delta T_{ad} \approx 2.8$ K and $\Delta S_m \approx 11$ Jkg$^{-1}$K$^{-1}$ can be achieved in a magnetic field change of $\Delta(\mu_0 H) = 1$ T [57, 58].

The third class includes materials with large thermal hysteresis like in Figure 2. One important example is the family of Heusler alloys with an inverse magnetostructural transition (see Figure 2(c,d)) ($\Delta T_{ad} \approx 3 - 4$ K, $\Delta S_m \approx 10 - 20$ Jkg$^{-1}$K$^{-1}$ in $\Delta(\mu_0 H) = 1$ T) [13, 59, 60]. Due to the large thermal hysteresis in Heusler alloys, high magnetic fields need to be applied in order to drive the transition fully reversible. When only small alternating magnetic fields of $\Delta(\mu_0 H) = 1 - 2$ T are available, their efficiency is drastically reduced. The same applies for Gd$_5$(Si,Ge)$_4$ with $\Delta T_{ad} \approx 3.5$ K and $\Delta S_m \approx 4 - 10$ Jkg$^{-1}$K$^{-1}$ in $\Delta(\mu_0 H) = 1$ T [61].

Among the magnetocaloric materials with first-order magnetic phase transitions, FeRh alloys hold a special place. It has been reported that the $\Delta T_{ad}$ in FeRh is 12.9 K in $\Delta(\mu_0 H) = 1.95$ T [31] or 7.9 K in $\Delta(\mu_0 H) = 1.93$ T [62], obtained in direct measurements, and these are the highest values ever recorded for any material under a magnetic field change up to 2 T. Although the raw material costs impede implementing FeRh in bulk form in a real magnetic refrigerator, this alloy is still interesting from a fundamental point of view as a model system for solid state refrigeration near room temperature. Our recent experiments show that under a magnetic field change of $\Delta(\mu_0 H) = 1$ T, the alloy Fe$_{49}$Rh$_{51}$ exhibits $\Delta T_{ad} \approx 3.6$ K and $\Delta S_m \approx 11$ Jkg$^{-1}$K$^{-1}$ [63].

## MCE and thermal hysteresis

One of the most important issues related to the magnetocaloric properties of materials like La(Fe,Co,Si)$_{13}$, La(Fe,Mn,Si)$_{13}$H$_x$ , Gd$_5$(Si,Ge)$_4$ , Heusler alloys and Mn-Fe-P-Si is the hysteresis of the first-order magnetostructural transition. The discontinuous nature of the transition is the feature that provides the large $\Delta S_m$. Only very recently, more research efforts have been directed to resolve issues related to transitional hysteresis [64, 65]. New ideas have been emerging how the hysteresis problem can be circumvented so that it can be possible to 'live with' or bypass hysteresis to some extent and how the hysteresis can be made

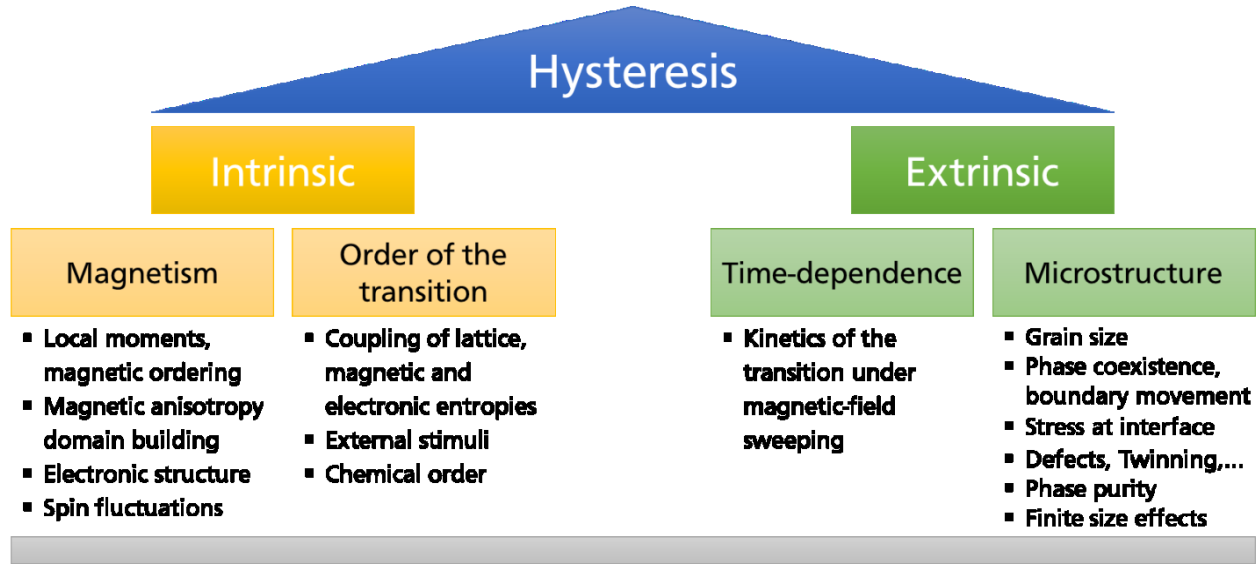

Figure 3: Hysteresis is related to intrinsic and extrinsic origins listed and grouped in the figure. Detailed investigations are needed for understanding how to overcome the hysteresis problem or how to obtain a substantial reversible magnetocaloric effect despite of hysteresis.

narrower by manipulating the microstructure or by tuning the composition [66, 67].

The temperature width of the transitional hysteresis is the main factor affecting the reversibility of $\Delta T_{ad}$ and $\Delta S_m$ when applying and removing a magnetic field. It is therefore necessary to understand the mechanisms causing hysteresis.

The origins of hysteresis can be separated into **intrinsic** contributions (linked to the electronic properties on the atomic scale) and into **extrinsic** influences (related to microstructure) as illustrated in Figure 3. The reversibility of the magnetocaloric effect, being essential for magnetic heat pumps, strongly depends on the width of the thermal hysteresis. Hysteresis can be tolerable in some cases, and in other cases, its width can be adjusted.

We distinguish two main categories for **intrinsic** origins: *magnetism* and the *order character of the transition* as shown in Figure 3. Our recent studies have shown that the magnetism is not only one of the key parameters which governs the magnetostructural transition, but also carries an influence on the width of the hysteresis [68]. A detailed understanding of the properties of the local moments, the magnetic ordering, magnetic anisotropy and domain formation within the framework of the complex electronic structure is required.

The order of the transition also plays a crucial role. While a first-order transition shows a hysteresis in general, a second-order transition is hysteresis-free. The big advantage of first-order materials is a relatively small magnetic field which is needed for the switching of the sample between the PM and FM states. This allows designing magnetocaloric cooling devices with a magnetic system made from permanent magnets with relatively small magnetic field changes of 1 - 2 T, in which it is possible to utilize the large entropy change associated with the magneto-structural transformation. On the other hand, the hysteresis drastically reduces the MCE when the magnetic field is applied under cycling conditions [63, 69, 70] and efforts are directed at overcoming the transitional hysteresis. The most important task here is to reach the tricritical point [71, 72] where a first-order transition becomes second-order. In this case, one can utilize the entropic benefits of the first-order transition without the reduction of the MCE under cycling. In particular, if the volume difference between the parent and the product phase can be eliminated, the hysteresis of a first-order transition narrows or even vanishes. If in addition, the crystallographic structures of the parent and product phases remain the same, the transition might become second-order and the hysteresis will be eliminated. In order to achieve this goal, good selection criteria for an identification of tricritical point are indispensable.

The order of the transition and the cross-over from first-order to second-order needs to be understood by theoretical and experimental methods. For example, the order of the transition and hysteresis width can be adjusted under the external stimuli, such as pressure [60], magnetic field, temperature and the local deviations in the chemical ordering and bonding [73].

The thermal hysteresis behavior can be manipulated by applying multiple stimuli other than magnetic field or by moving in minor loops of the transition instead of completely transforming the material from one phase to the other. As demonstrated in $Gd_5Si_2Ge_2$ [74] and $La(Fe,Si)_{13}H_x$ [75], these conventional MCE first-order type materials show an inverse barocaloric effect. Referring to the Clausius-Clapeyron equation for the application of hydrostatic pressure, the magnitude of the barocaloric effect is related to the volume change during the transformation and the shift of the transition temperature by pressure. In $Ni_{49.26}Mn_{36.08}In_{14.66}$ an external pressure shifts the equilibrium martensitic transformation temperature $T_t$ by 2 K $kbar^{-1}$ [76], while for the $Ni_{45.2}Mn_{36.7}In_{13}Co_{5.1}$ alloy this increase of the $T_t$ is with 4.4 K $kbar^{-1}$ much more pronounced (Figure 4). So it is also important to state that inverse MCE materials like Ni-Mn-Co-In demonstrate a conventional barocaloric effect. As shown in Figure 4, the magnetic hysteresis can be significantly reduced if the sample is magnetized without bias stress but demagnetized under an external pressure. Furthermore, theoretical calculations predict that the efficiency of the magnetocaloric material can be improved when implemented in a device with precisely adjustable applied magnetic field and pressure, compared to the device where only the magnetic field can be varied [77].

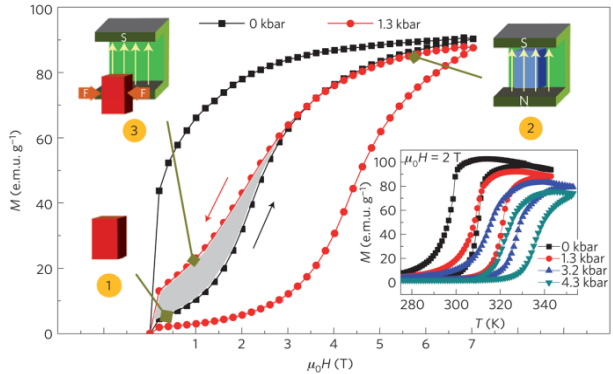

**Figure 4: The large thermal irreversibility can be overcome by the combination of applied magnetic and mechanical forces. M-T curves under 0 and 1.3 kbar hydrostatic pressure at 308 K are shown for $Ni_{45.2}Mn_{36.7}In_{13}Co_{5.1}$. The bottom right corner insert shows the shift of martensitic transition temperatures by the application of a hydrostatic pressure up to 4.3 kbar. The forward and reverse transitions can be induced in a relatively low field (with little hysteresis—shaded region in the main figure) when the sample is magnetized in zero pressure but demagnetized under an external pressure. This process is demonstrated schematically as well. (originally published in [60])**

The **extrinsic** origins of hysteresis can also be divided into two categories, as *microstructural* and *time-dependent* as shown in Figure 3. The microstructure involves grain-size, phase coexistence and phase-boundary movement, interfacial stress, defects and phase purity. Besides this, the size of the sample can also affect the thermal hysteresis, which acts mainly on the mesoscopic scale. The time dependence of these transformation mechanisms is an issue on its own and involves the response to external stimuli. Our recent experiments show that the hysteresis increases when a magnetic field is applied with very high change rates [78, 79]. Since first-order transitions are driven by nucleation and growth, these processes need a certain time to take place. When the magnetic-field rate is too high, the material can simply not follow and consequently a broadening of the hysteresis is observed. These limitations in the transformation speed need to be considered especially for magnetocaloric applications operating at high frequencies.

An additional problem of materials undergoing a sharp first-order transition is the volume change leading to mechanical stress causing, and, in turn, the occurrence of dislocations, micro-cracks, and even major cracks resulting in the pulverization when the material undergoes several temperature or field cycles [80]. The smaller fragments require less elastic energy to complete the transformation and consequently a reduced hysteresis is observed [81, 82].

Investigations of the first-order transition kinetics in the giant magnetocaloric material $LaFe_{11.8}Si_{1.2}$ were performed by an in-situ X-ray diffraction experiment and by magnetometry. Due to the difference of the lattice parameters of the ferromagnetic and the paramagnetic phase, they mechanically interact with each other in the temperature range of coexistence. As a result, huge strains appear at the phase boundary which leads to the formation of cracks during the first cooling sequence. The corresponding X-ray tomography image of the sample is shown in Figure 5. One can see that the interlocked particles are separated from each other by significant gaps of widths in the range $d$ = 2–6 µm, but the whole particle ensemble nonetheless holds together despite the existence of cracks. Magnetic measurements were done on (1) the sample with cracks as a whole and (2) all separate fragments together, obtained after the sample was disassembled into individual grains (Figure 5).

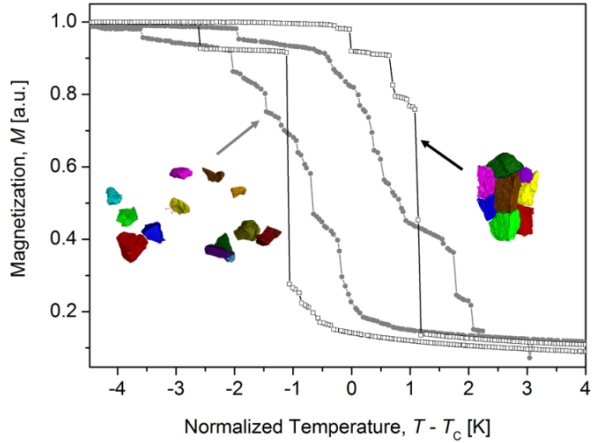

**Figure 5:** The magnetic transition of an interlocked particle ensemble (open squares) is very sharp even though most particles are separated by cracks. For a well separated particle ensemble (solid circles), the transition broadens significantly, indicating that proximity of particles – and not necessarily a mechanical connection – is required to maintain a sharp, first-order like transition. The sample images show the measured particle ensembles imaged with computed tomography [105]. (copyright granted, License Number: 3847701318380)

Isofield magnetization curves of this interlocked structure in an external magnetic field of $\mu_0 H = 10$ mT demonstrate a very sharp transition (see squares in Figure 5), indicating that the interlocked particle ensemble still transforms together as one piece. As the tomography reveals that all particles of the interlocked ensemble are connected to the surface and thus can change their volume without the severe constraints that a particle would experience when surrounded by other particles. The magnetic transition of clearly separated particles (Figure 5, solid circles) is much broader compared to the interlocked particle ensemble. The transformation proceeds in many small steps presumably representing the $T_t$ distribution of the individual particles. This raises the question, which coupling mechanism is causing the sharp transition in the interlocked state, where particles are only loosely connected. Following [83] we propose that magnetostatic interactions, which do not require direct contact between the particles, facilitate the nucleation in neighboring particles and lead to an avalanche-like sharp transition.

In addition to this effect, depending on the choice of the measurement protocol, different values of the MCE might be observed in materials with a first-order transformation being associated with thermal hysteresis. Most of the $\Delta S_m$ data reported for materials with first-order transitions are obtained under a "discontinuous" measurement protocol [84, 85], which allows maximizing the $\Delta S_m$ value, but at the same time it is being far from real operational conditions. A magnetic refrigerator operates in a cyclic manner, which is usually not replicated in laboratory tests. Whereas single sweep laboratory characterization provides important information on the material properties, these measurements are not necessarily adequate for an assessment of the suitability of a material for its application for room temperature magnetic refrigeration [70].

In our recent works [63, 86], a rigorous method is proposed for extracting the cyclic magnetic entropy change of a magnetocaloric material even if the characterization is performed under quasistatic noncyclic conditions. Results show that the cyclic response coincides with the intersection of the heating and cooling curves measured using protocols, which prevents the appearance of spurious spikes upon the application of the Maxwell relation. This method provides a basis for a comparison of the suitability of different hysteretic magnetocaloric materials for their application in a magnetic refrigerator.

## La(Fe,Co,Si)$_{13}$

In the La-Fe-Si material family, the thermal hysteresis can be tuned down by the specific substitution of elements [87, 88]. The greater challenge is to increase the transformation temperature of the itinerant electronic transition to room temperature without diminishing the magnetocaloric properties.

To show the whole picture of magnetocaloric properties of the La(Fe,Co,Si)$_{13}$ alloys, we collected our $\Delta S_m(T)$ and $\Delta T_{ad}(T)$ dependences in a single plot in Figure 6(a). The substitution of Co results in a shift of the Curie temperature up to room temperature, but it is simultaneously accompanied by a decrease of $\Delta S_m$ and $\Delta T_{ad}$. Increasing the Co content changes the type of the transition from first-order to second-order. Consequently, the magnetovolume change shows the same trend as $\Delta T_{ad}$. This demonstrates that the magneto-elastic contribution plays a key role in order to obtain a large MCE but at the same time, the thermal hysteresis grows.

In our previous work [23], we demonstrated how more sophisticated geometries can be made by using selective laser melting, a rapid prototyping technique to form three-dimensional shapes. Using the excellent magnetocaloric material La(Fe,Co,Si)$_{13}$ as starting-powder, we made two geometries: a wavy-channel block with a high surface-to-volume ratio and an array of fine-shaped rods, which eliminate unwanted heat conduction along the magnetic part. After annealing treatment, the geometries were intact and survived more than $10^6$ cycles of applying a magnetic field while still maintaining good magnetocaloric properties. This fabrication approach is promising for making near-net shaped magnetic refrigerants with superior heat transfer properties and performance.

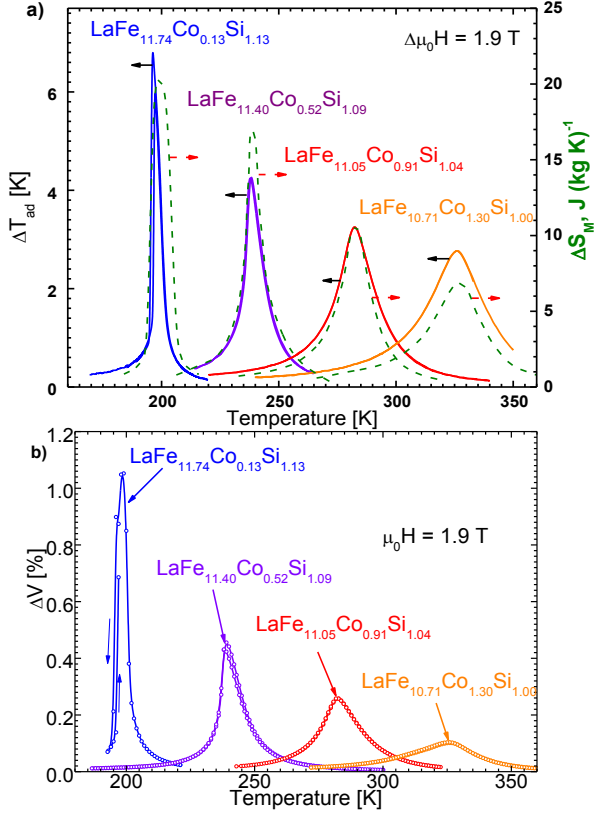

**Figure 6:** (a) Isothermal magnetic entropy change $\Delta S_m$ and adiabatic temperature change $\Delta T_{ad}$ for 4 selected La(FeCoSi)$_{13}$ compounds measured under magnetic field change $\Delta\mu_0 H$=1.9 T, (b) Isothermal magnetovolume effect.

## La(Fe,Mn,Si)$_{13}$H$_X$

By modulating the magnetic environment of the Fe-atoms, the ferromagnetic state can be stabilized either by elemental substitution of Fe (by Co, Mn, Si, etc.) or interstitial insertion of small atoms like C, N, B and H, hence, $T_t$ can be changed from 190 K up to room temperature and above (340 K). The hydrogenation of La(Fe,Si)$_{13}$ is the most efficient method to bring $T_t$ near room temperature since the large magnetocaloric effect is retained. On the other hand, in order to precisely adjust $T_t$ to the desired working temperature, partial hydrogen absorption or desorption is required. Process parameters like pressure, temperature and dwell need to be adjusted carefully. Moreover, the long-term stability of such partial hydrogenated compounds is not guaranteed as has been reported recently [89]. Therefore, a quaternary compound with Mn has been suggested to adjust $T_t$ of the fully hydrogenated material. Comparable to Si and Co, with increasing Mn content, the type of the magnetic transition is shifted gradually from first- to second-order and the MCE decreases.

We and others demonstrated that the hydrogenation of La(Fe,Si,Mn)$_{13}$ is a simple and efficient way to utilize this compound as a first-order type material at room temperature [90, 91]. The $T_t$ of the non-hydrogenated sample is determined by the Mn content of the parent sample, as well as its MCE. This is attributed to the gradual change of the phase transition from first- to second-order and the reduction of the thermal hysteresis of the non-hydrogenated sample as one can see in Figure 7 (a). In Figure 7 (b), the magnetocaloric properties of fully hydrogenated non-compacted powder with working temperatures around room temperature are shown. These materials were commercially produced by Vacuumschmelze GmbH. Like in the case of the non-hydrogenated compound, Mn decreases the transition temperature and the MCE. At the same time, the transition is of first-order type and the thermal hysteresis is negligible.

Due to the hydrogenation process, the decrepitation of the material into powder is inevitable. Therefore, it is not possible to manufacture heat exchangers as plates in bulk form. A simple solution is the compaction of powder into thin plates. Adhesive-bonding techniques can provide mechanical stability, corrosion protection and enable the production of net shaped modules in a single step manufacturing process. In order to obtain an excellent magnetocaloric heat exchanger from a precursor material as hydrogenated La(Fe,Si,Mn)$_{13}$, a systematic study on making an epoxy-composite with this material was performed [55] and the influence of the powder particle size, adhesive type, adhesive concentration and compaction pressure on the magnetocaloric properties of the polymer-bonded La(Fe,Mn,Si)$_{13}$H$_x$ material was investigated in detail.

After the optimization of all compaction parameters, the adiabatic temperature change (red curve in Figure 7 (b)) of the fully-dense La(Fe,Mn,Si)$_{13}$H$_x$ composite in a magnetic field change of $\Delta(\mu_0 H)$ = 1.93 T (4.9 K) is comparable to the $\Delta T_{ad}$ of Gd metal ($\Delta T_{ad}$ = 5.1 K) and significantly exceeds the $\Delta T_{ad}$ of La(Fe,Co,Si)$_{13}$ ($\Delta T_{ad}$ = 3.1 K). The value of the volumetric magnetic entropy change of the composite La(Fe,Mn,Si)$_{13}$H$_x$ (63 mJcm$^{-3}$ K$^{-1}$) is comparable to the one of bulk La(Fe,Co,Si)$_{13}$ (70 mJcm$^{-3}$ K$^{-1}$) and 50% higher than the one of Gd metal (40 mJ cm$^{-3}$ K$^{-1}$). The thermal conductivity of the polymer-bonded La(Fe,Mn,Si)$_{13}$H$_x$ is about 5 WK$^{-1}$m$^{-1}$ near room-temperature [55]. This is 2.5 times less than the one of bulk La(Fe,Co,Si)$_{13}$ and Gd metal, but still 5 times higher than that of MnAs and Fe$_2$P-type materials. This means that a device using a composite material will have 1.5 times lower maximum operation frequency than a device using bulk material.

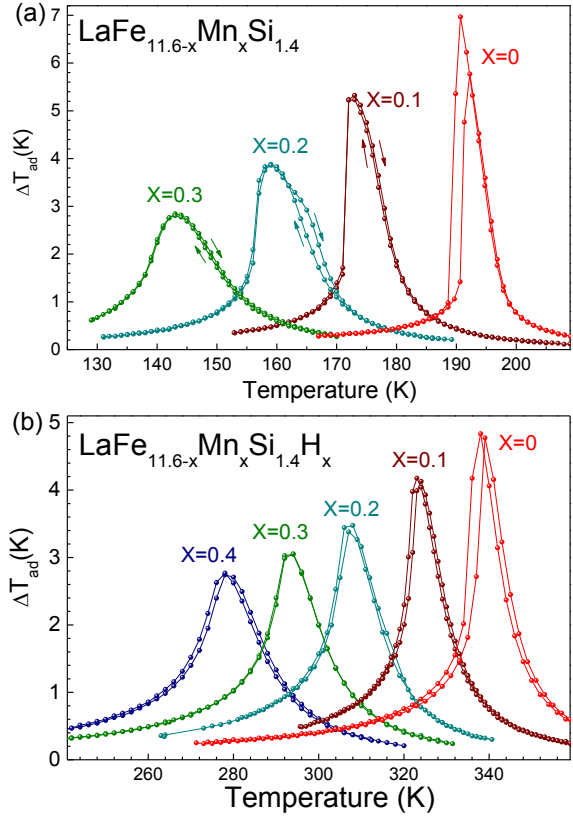

**Figure 7:** Adiabatic temperature change $\Delta T_{ad}$ as function of temperature for parent compounds LaFe$_{11.6-x}$Mn$_x$Si$_{1.4}$ (a) and their hydrogenated samples (b). The Mn content increases from right to left, i.e. with decreasing $T_{Peak}$.

## Fe$_2$P-type compounds

Another class of very promising rare-earth-free materials are the Fe$_2$P-type compounds. The family is derived from the prototypical Fe$_2$P-base compound, showing a sharp first-order phase transition at $T_t$ = 217 K [92, 93]. In 2002 the compound MnFeP$_{0.4}$As$_{0.55}$ was reported by Tegus et al. exhibiting a giant magnetocaloric effect near room temperature at which the structural and magnetic transformations coincide [94]. Besides the large thermal hysteresis, the other drawback of this type of alloys is the content of toxic Arsenic, limiting the use in a domestic refrigeration appliance. Since the discovery of this alloy, large efforts have been focussed on tuning the transition and especially on the substitution of critical elements in this system. The As could be replaced by Si completely, conserving the magnetocaloric properties. Furthermore, the transition temperature is highly tuneable by adjusting the Mn and Si content [95]. In this material class, the first-order transition also manifests itself in a large volume change of the hexagonal structure. Conventionally, the desired phase is prepared by ball milling of precursor powders, compaction and a subsequent heat treatment procedure for sintering [95] or by melt spinning [96].

Studies in the class of Mn-Fe-P-Si-type alloys have shown that the hysteresis could be minimized by fine adjustment of the Mn/Fe and P/Si ratio. The drawback of this adjustment is the decrease in overall magnetization and therefore a decrease in $\Delta M$ at the transition. In order to increase both values, one has to work at the optimum composition of MnFeP$_{0.66}$Si$_{0.33}$ [97, 98]. This composition has a low value of $\Delta T_{ad}$ (≈ 2 K in 1 T), but at the same time, shows a rather high value of $\Delta S_m$ = -8 Jkg$^{-1}$K$^{-1}$ [99]. A major drawback of these compounds is mechanical instability. The instability is considered to be from the volume change of +0.2 % [97] when crossing the transition temperature leading to embrittlement of the bulk samples.

Recently above mentioned issues were tackled by applying internal pressure to the compound by adding Boron to the system [100]. For the compound MnFe$_{0.95}$P$_{0.595}$B$_{0.075}$Si$_{0.33}$ a cyclic adiabatic temperature change of $\Delta T_{ad}$ = 2.55 K with a cyclic $\Delta S_m$ ≈ 10 Jkg$^{-1}$K$^{-1}$ in a field change of 1 T with a thermal hysteresis of 1.6 K was observed. This enhancement of properties is ascribed to the larger field dependence of $T_C$ and lower latent heat of this material compared to Mn-Fe-P-Si. Furthermore, the mechanical integrity of this compound was reported to be optimized to a point where no breaking or cracking of the samples could be observed, even after 10000 cycles of magnetization and demagnetization [101]. This improvement of the structural integrity was ascribed to the fact that there is no discontinuous volume change present when passing through the transition but still preserving a comparable change in a/c lattice parameters to the base Mn-Fe-P-Si compound.

It was also shown that the B containing compounds Mn$_x$Fe$_{(1.95-x)}$P$_{(1-y-z)}$Si$_z$B$_y$ are highly tuneable while preserving the giant magnetocaloric effect with a limited thermal hysteresis in broad temperature range [58]. Very recent works show that Boron can also be substituted by Nitrogen, possibly increasing the industrial applicability of this compound [102].

In summary, the compounds of Mn-Fe-P-Si-B/N are, as the La(Fe,Mn,Si)$_{13}$H$_x$-type alloys, a highly promising material class for magnetic refrigeration due to their precisely tuneable transition temperatures with low hysteresis, high reversible values of $\Delta S_m$ and $\Delta T_{ad}$ at low fields applicable in a refrigeration device. Furthermore, they also consist of cheap, nontoxic and readily available elements.

## Heusler alloys

The large magnetocaloric effect in Heusler compounds is due to the first-order magnetostructural transition between the low-temperature martensite and the high-temperature austenite phase [60, 103]. Significant entropy and temperature changes can be achieved with these materials [104]. On the other hand, the thermal hysteresis is large which impedes the cyclability of the effect. Direct measurements on the Heusler alloy Ni$_{45.7}$Mn$_{36.6}$In$_{13.5}$Co$_{4.2}$ revealed a very large adiabatic temperature change $\Delta T_{ad}$ of -8 K in a magnetic field change of only 1.93 T as shown in Figure 8 (a). Because of the large thermal hysteresis, which is in the range of 10 K, this large effect cannot be achieved in the second field application cycle.

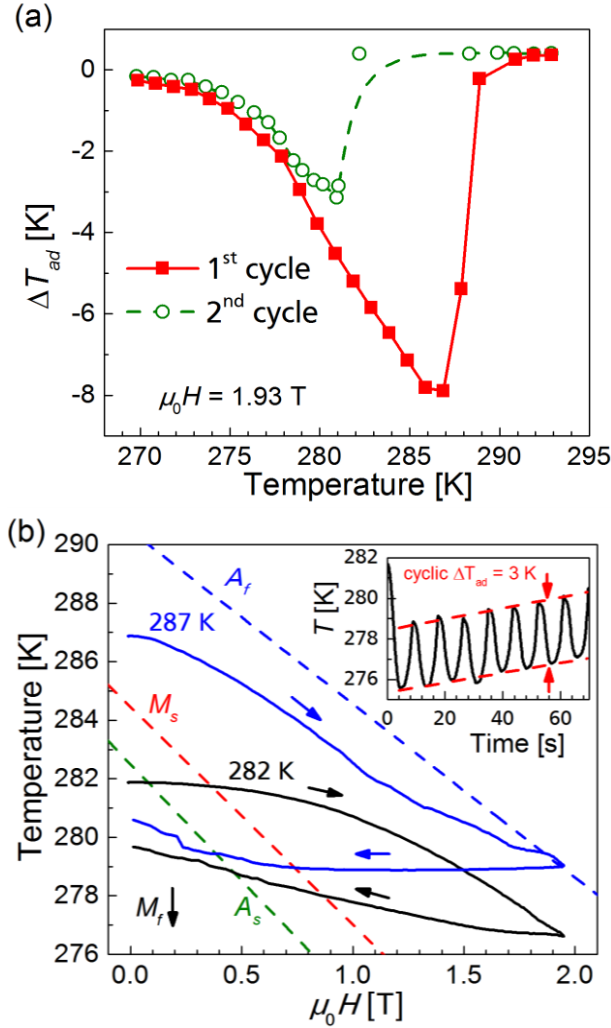

Figure 8: (a) $\Delta T_{ad}$ upon first and second application of 1.93 T. (b) Temperature progression of the Heusler compound $Ni_{45.7}Mn_{36.6}In_{13.5}Co_{4.2}$ starting from different temperatures. The inset shows the cyclic response of the material. [3] (copyright granted, license Number: 3850160649979)

The negative effect of thermal hysteresis can be reduced in Heusler alloys when following a minor loop of hysteresis instead of completely transforming the material between martensite and austenite [59]. Such a minor loop is in principle only possible if the transition is continuous and not jump-like as shown schematically in Figure 2(d). However, real magnetocaloric Heusler alloys have always a finite transition width. This was explained by partly preventing the nucleation of martensite in the mixed-phase region of minor loops, which would require more energy than the simple phase boundary movement. We related this effect to the strong sensitivity of the transition temperature $T_t$ to internal stress, being created by the large volume change of the transition.

The reversible magnetocaloric effect accounts to -3 K when moving in minor loops of hysteresis as shown in Figure 8 (b). The origin of this was investigated by temperature-dependent optical microscopy and it turned out that the energy-intense nucleation of the low-temperature phase can be partly avoided in the minor loop. In this case, the sample transforms mainly by phase-boundary movement, which is less costly in energy, and therefore, the reversibility of the first-order transition is increased [59]. This important finding brings Heusler alloys closer to application, since the reversible $\Delta T_{ad}$ is now comparable to other magnetocaloric materials such as $La(Fe,Si,Co)_{13}$.

## Conclusions

In this paper, we have summarized the main causes of hysteresis in different magnetocaloric material classes and discussed possibilities to reduce it or how to utilize a material despite a large hysteresis. We separated the origins for hysteresis into intrinsic contributions (linked to the electronic properties on the atomic scale) and extrinsic factors (related to microstructure). The reversibility of the magnetocaloric effect, essential for magnetic refrigeration, strongly depends on the width of the thermal hysteresis. It is therefore absolutely required to measure the adiabatic temperature change as well as the isothermal entropy change under cyclic conditions in order to assess the suitability of a material in a meaningful way. The width of the hysteresis reduces when approaching the tricritical point at which the mismatch between the high and the low temperature phase vanishes. This can be reached for instance by chemical substitution or by the application of external stimuli. Even though the reversibility enhances by this approach, a reduction of the magnetocaloric properties can occur when the order of the transformation changes. In contrast, a material with a first-order transition and a certain hysteresis should have a sharp transformation in order to obtain large MCE. Such a transition is driven by nucleation and growth, therefore the size of the material as well as the kinetics may also influence the hysteresis.

We reviewed recent achievements for the most promising magnetocaloric materials with great potential for room temperature applications. Non-hysteretic materials such as prototypical Gd act as a reference to see whether hysteretic materials can be tuned to outperform them.
By adding Co to the material $La(Fe,Si)_{13}$ it is possible to shift $T_t$ to room temperature. Furthermore, the character of the phase transition of the material changes from first- to second-order and thus the thermal hysteresis is reduced. However, this effect is accompanied by a reduction of the MCE.
Another possibility to tune $La(Fe,Si)_{13}$ is the hydrogenation of the base material. By this approach, the phase transition is still of first-order type and the MCE remains large. At the same time, the hysteresis remains small enough to make the material very promising. The main drawback is the high brittleness and poor machinability, which can be overcome by binding the material in a matrix. This however dilutes the MCE.

$Fe_2P$-type compounds have been successfully tuned to exhibit a transformation near room temperature. Reports shoiw that the addition of B and N can largely eliminate the volume change in the material and by this the hysteresis could be reduced down to 1 K. Furthermore, the addition of B prevents cracking and the material could prove to be useful for application.

Heusler alloys can also be useful despite a large hysteresis. By operating in minor loops of the thermal hysteresis, it is still possible to have a significant magnetocaloric performance under cyclic conditions. The reason for this is the large number of martensitic nucleation sites that remain in the material if it is not transformed completely.

In materials with a large volume change during the transformation, the application of pressure influences the transition temperature and the width of the thermal hysteresis. When being close to the tricritical point, the thermal hysteresis can be reduced significantly. If this condition is not fulfilled, the combination of both an external magnetic field and pressure can help to enhance the cyclability.

In the last decade, a significant progress in the optimization of known materials and in the identification of new potential materials emerged. However, these achievements could not yet drive this interesting field of magnetic refrigeration to market entry on a large scale. It is obvious that a magnetocaloric material with a drastically enhanced reversible entropy and temperature change would draw a lot attention in fundamental material science and physics, but this alone is not sufficient to design a solid-state refrigerator being competitive with conventional technology in price, weight and efficiency. In order to achieve this goal, considerable joint efforts of fundamental science on the one side and mechanical engineering on the other side need to be made. This is the indispensable prerequisite to pave the way for a bright future of magnetic refrigeration.

**Acknowledgments**

This work has been in part funded by the European Community's 7th framework programme under grant agreement 310748 "DRREAM" and also the DFG SPP 1599 priority programme. We also want to thank the DFG for financial support in the framework of the Excellence Initiative, Darmstadt Graduate School of Excellence Energy Science and Engineering (GSC 1070).